\documentclass[twocolumn,prl]{revtex4}

\ifx\pdfoutput\undefined
  \usepackage[dvips]{graphicx}
\else
  \usepackage[pdftex]{graphicx}
\fi
\usepackage{dcolumn}
\usepackage{amsmath}
\usepackage{latexsym}
\begin{document}
\title[Josephson Tunnel Junctions with Nonlinear
Damping]{Josephson Tunnel Junctions with Nonlinear
Damping \\ for RSFQ-Qubit Circuit Applications}

\author{A.\,B.~Zorin}
\author{M.\,I.~Khabipov}
\author{D.\,V.~Balashov}
\author{R.~Dolata}
\author{F.-I.~Buchholz}
\author{J.~Niemeyer}

\affiliation{Physikalisch-Technische Bundesanstalt, Bundesallee
100, 38116 Braunschweig, Germany}%
\begin{abstract}
We demonstrate that shunting of Superconductor-Insulator-Superconductor (S-I-S)
Josephson junctions by Superconductor-Insulator-Normal metal (S-I-N) structures
having pronounced non-linear $I$-$V$ characteristics can remarkably modify the
Josephson dynamics. In the regime of Josephson generation the phase behaves as
an overdamped coordinate, while in the superconducting state the damping and
current noise are strikingly small, that is vitally important for application
of such junctions for readout and control of Josephson qubits. Superconducting
Nb/AlO${_x}$/Nb junction shunted by Nb/AlO${_x}$/AuPd junction of S-I-N type
was fabricated and, in agreement with our model, exhibited non-hysteretic
$I$-$V$ characteristics at temperatures down to at least $1.4\,$K.

\end{abstract}
\maketitle

The \emph{overdamped} Josephson junctions are the key elements of the SQUIDs
and Rapid Single-Flux-Quantum (RSFQ) logic circuits \cite{Likh-Sem,Bunyk}. Due
to non-hysteretic (single-valued) $I$-$V$ characteristics, they allow the
convenient readout of critical currents of interferometers and the generation
and processing of Single Flux Quantum (SFQ) voltage pulses associated with
2$\pi$ slips of the Josephson phase. Recently, RSFQ electronics have been
considered as possible complementary digital electronics for serving the
Josephson qubits. Specifically, these generic superconducting circuits should
allow the efficient control, readout and processing of information in terms of
the SFQ pulses (see, for example, Refs. \cite{Sem-Aver,Crankshaw}). Among the
challenging problems of matching RSFQ and qubit circuits, the principal one
seems to be the back action of RSFQ circuits on qubit \cite{Sem-Aver,Wulf}.
This back action is due to current noise which stems from unfavorably large
damping in the RSFQ-circuit junctions. Moreover, this noise is essential even
in the quiescent (zero-voltage) state of the junctions. By means of direct
inclusion (or inductive coupling) of an RSFQ-circuit input in (to) a qubit
based on underdamped Josephson junctions, the current (or flux) noise may cause
dramatic decoherence of the qubit: pure dephasing due to low-frequency
components of the noise and both dephasing and relaxation due to the frequency
components close to the transition frequency of the qubit $\nu_{01}$
\cite{Makhlin}. (Typically, $\nu_{01}=$ 10-30~GHz $\ll \Delta/h$, $\Delta$
being the superconductor energy gap.) Then quantum manipulation of the qubit is
impossible.

At an operating temperature $T \lesssim 0.5\,T_c$ ($T_c$ being the critical
temperature of the superconductor) the most technological Josephson junctions
of the Superconductor-Insulator-Superconductor (S-I-S) type of low and moderate
critical current density $j_c$ are \emph{underdamped} \cite{Barone}. In these
junctions, the transient dynamics of the Josephson phase $\phi$ prevents their
immediate use for generating SFQ pulses. When the bias current $I$ rises above
the critical value $I_c$, the junction switches from the superconducting to the
resistive state, while reset (return to the superconducting state) occurs at
the much smaller current $I_R\ll I_c$. Overdamped behavior is usually achieved
by shunting the junctions with a normal resistance of sufficiently small $R$.
In this case, the dynamics is described by the Resistively Shunted Junction
(RSJ) model \cite{McCum,Stew} and the condition of sufficient damping is
formulated as $\beta_c\lesssim 2$, where the Stewart-McCumber parameter is
given by
\begin{equation} \label{beta_c}\beta_c = (2\pi /\Phi_0)I_c R^2 C,
\end{equation}
$\Phi_0=h/2e$, flux quantum and $C$, junction capacitance. For such $\beta_c$,
the hysteresis in autonomous junctions is small, $I_R/I_c \gtrsim 80\%$, which
ensures correct operation of an RSFQ circuit. The real part of the junction
zero-voltage admittance is, however, large, Re$Y(\omega)=G\equiv R^{-1}$. The
corresponding spectral density of current fluctuations,
\begin{equation} \label{S_I}S_I(\omega)=(\hbar\omega
G/\pi)\coth(\hbar\omega/2k_BT),
\end{equation}
is also large at the qubit frequencies, i.e.,
$\omega/2\pi\lesssim \nu_{01}$.

One of the possible ways to suppress the current fluctuations consists in using
a very transparent tunnel barrier having a specific tunnel resistance $\rho\sim
1\,\Omega\cdot\mu\textrm{m}^{2}$ yielding critical current density
$j_c=10^5-10^6\,\textrm{A/cm}^2$ \cite{JN,Miller,Patel}. In the dynamic state
($V\neq 0$), the effective damping dramatically increases and the return
current $I_R$ can approach $I_c$ \cite{Harris,Z-L_FNT,Z-L-T}. At $T\rightarrow
0$ and $\omega<2\Delta/\hbar$, the zero-bias admittance Re$Y(\omega)=
\sigma_{\textrm{qp}}(\omega) + \sigma_{\textrm{p-qp}}(\omega) \cos \phi$ (where
$\sigma_{\textrm{qp}}(eV/\hbar)$ is quasiparticle and
$\sigma_{\textrm{p-qp}}(eV/\hbar)$ is pair-quasiparticle-interference
conductances \cite{Jos}), is remarkably small (for the BCS density of states,
Re$Y\propto \exp(-\Delta/k_BT)$  \cite{Wert,LarOvch}). So one could expect low
current noise in this frequency range. Unfortunately, the barriers of these
junctions often have pin-hole defects leading to large $\sigma_{\textrm{qp}}$
at small $V$ \cite{Patel}. Therefore, these junctions can hardly be implemented
in multi-junction circuits and guarantee small noise.

In this Letter we frame a concept of \emph{nonlinear} shunts which can be
realized by means of Superconductor-Insulator-Normal metal (S-I-N) junctions.
It is well known that the $I$-$V$ curve of an S-I-N junction is strongly
non-linear. For the BCS superconductor, it is given by
\begin{equation} \label{I_SIN}
I_{\textrm{SIN}}(V) = \frac{1}{eR} \int^{\infty}_{\Delta} dE
\frac{[f(E-eV)-f(E+eV)] E }{(E^{2}-\Delta^{2})^{1/2}},
\end{equation}
$R$ being the junction resistance at large voltage ($eV \gg \Delta$) and $f(E)
= 1/[1+\exp (E/k_B T)$], the Fermi function. At low temperature,
$\delta=\Delta/k_BT \gg 1$, the small-bias current
$I_{\textrm{SIN}}(V\ll\Delta/e)$ and the zero-bias conductance $G_0$ of the
junction are small, $G_0R= (2\pi\delta)^{1/2}\exp(-\delta)\ll 1$
\cite{Tinkham}. Therefore, one can expect only small current noise to be
generated by this element at $\omega \ll \Delta/\hbar$ (see Eq.\,(2) in which
$G$ should be substituted for $eI_{\textrm{SIN}}(\hbar\omega/e)/\hbar\omega$
\cite{DahmRogovin}). The only question which arises is whether it is possible
to ensure sufficient damping in a Josephson junction by shunting it with an
S-I-N junction.

The answer is apparently positive, first of all, because the dynamics of S-I-S
junctions with very small quasiparticle leakage current can theoretically be
made overdamped by increasing $j_c$. Quantitatively, the condition can be
formulated in terms of $\beta_c \lesssim 0.1$, assuming that $R$ in Eq.\,(1) is
the tunneling resistance of the S-I-S junction at $V\gg 2\Delta/e$ (see
calculations made within the framework of the microscopic model of Josephson
tunneling \cite{Wert,LarOvch} in Ref.\,\cite{Z-L-T}). The simpler model of
Prober et al. \cite{Prober}, approximating the quasiparticle curve by three
straight line segments, also gives the values of $\beta_c$ sufficient for
suppressing the hysteresis in the range of $0.1-1$.

\begin{figure}[t]
\begin{center}
\leavevmode
\includegraphics[width=2.8in]{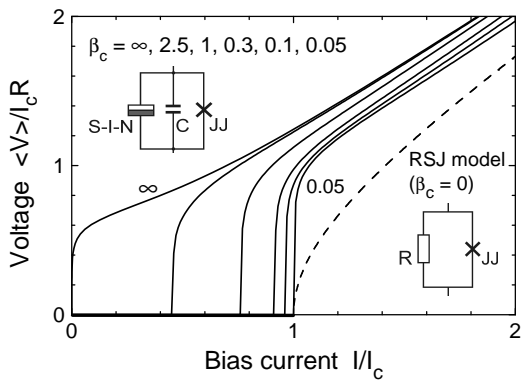}
\caption {$I$-$V$ curves of S-I-N shunted Josephson junction calculated for
different values of parameter $\beta_c$ with the bias current sweep starting
from the large values (solid curves). The values of the other parameters are:
$\delta\equiv\Delta/k_BT=10$ and $\alpha\equiv eI_cR/\Delta=\pi/2$. The dashed
line is the hyperbolic-shaped $I$-$V$ curve given by the RSJ model with zero
capacitance \cite{McCum,Stew}.} \label{Scheme}
\end{center}
\end{figure}

We carried out simple simulations of a Josephson junction shunted by an S-I-N
junction. In this model we assumed that the total current $I$ from a source
branches into the capacitive, dissipative and Josephson components:
\begin{equation} \label{motion}
C\frac{dV}{dt} + I_{\textrm{SIN}}(V) + I_c\sin \phi= I,
\end{equation}
where $C$ is the total capacitance of both junctions. Current
$I_{\textrm{SIN}}(V)$ is given by the BCS dependence Eq.\,(3) with the instant
voltage $V$ as an argument. The Josephson current amplitude $I_c$ was assumed
to be frequency-independent and the quasiparticle and pair-quasiparticle
interference components of current through the S-I-S junction were omitted.
These assumptions are reasonable because in the interesting voltage range (up
to approximately the characteristic value $V_c\equiv I_cR$), the damping should
be produced by the S-I-N junction. Finally, we omitted the ac terms in the
current through the S-I-N junction, related to the imaginary part of its
impedance and leading to an equation of motion of the integral type. The
estimation had, however, shown that taking these terms in account did not lead
to an essential correction of the dc $I$-$V$ curves, but significantly
complicated the calculations.

\begin{figure}[t]
\begin{center}
\leavevmode
\includegraphics[width=2.6in]{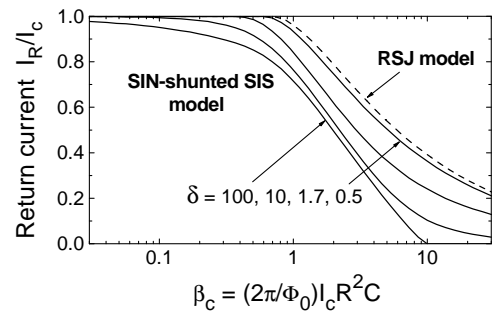}
\caption {Return current versus parameter $\beta_c$ calculated in our model for
different values of $\delta$ and fixed $\alpha = \pi/2$ (solid lines) and RSJ
model (dashed line) \cite{McCum}.} \label{IV-curve}
\end{center}
\end{figure}

We numerically solved Eq.\,(\ref{motion}) for different values of parameter
$\delta$.  We found out that in the most interesting case of pronounced
nonlinearity, i.e. large $\delta$ $(\gtrsim 10)$, sufficient damping can be
achieved by reasonable reduction of $\beta_c$ only if the parameter $\alpha
\equiv eV_c/\Delta \gtrsim 1$. Figure\,1 shows the effect of $\beta_c$ on the
$I$-$V$ curves for fixed $\delta=10$ (yielding the ratio of the zero-bias and
large-bias conductances $G_0R\approx3\times 10^{-4}$) and $\alpha=\pi/2$. The
leftmost curve which corresponds to $\beta_c=\infty$ is merely the $I$-$V$
curve of the S-I-N junction for the given $\delta$. At $\beta_c <1$, hysteresis
is sufficiently suppressed. The shape of these curves has a much steeper
voltage rise at $I=I_c$ than that given by the RSJ model (dashed line). This is
the result of deficient low-frequency damping which the system compensates by
large damping at high frequencies, corresponding to large mean voltage,
$\langle V\rangle\geq \Delta/e$. The return current as function of $\beta_c$
(see Fig.\,2) generally mimics the dependencies given by the RSJ model,
especially for small $\delta$ (cf. dashed line), as well as the
piecewise-resistor model \cite{Prober} and the microscopic model \cite{Z-L-T}
(both not shown).

\begin{figure}[t]
\begin{center}
\leavevmode
\includegraphics[width=3.2in]{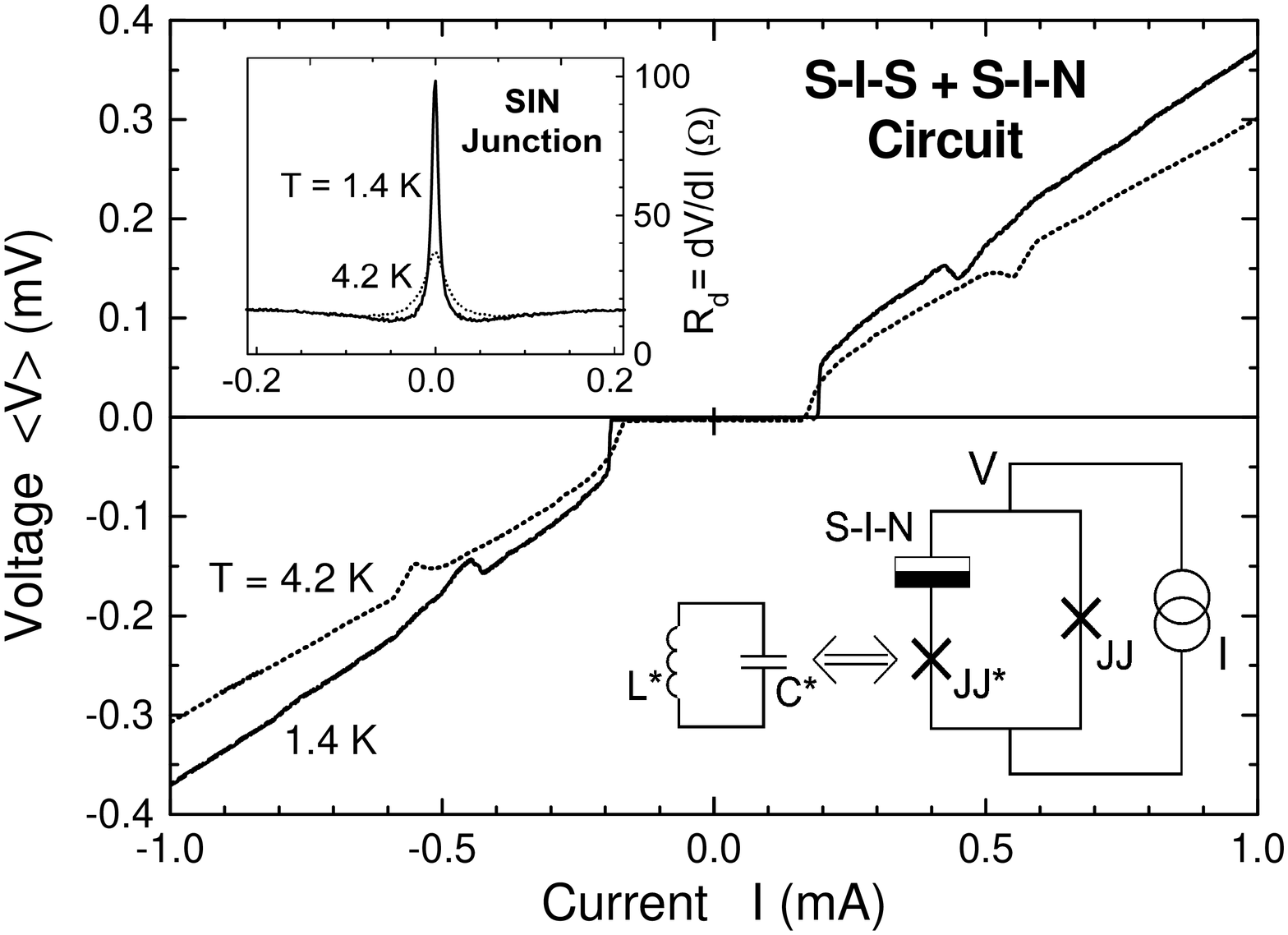}
\caption {Non-hysteretic $I$-$V$ curves of S-I-N-shunted SIS junction measured
at 4.2\,K (dotted line) and 1.4\,K (solid line). Upper inset shows the
$dV/dI$-$I$ curves of a stand-alone S-I-N junction manufactured from the same
sandwich. Lower inset shows the electric diagram of the circuit including an
additional S-I-S junction JJ* which was formed beneath the S-I-N structure
because of merely technological simplicity.} \label{IV-curve}
\end{center}
\end{figure}

To implement the idea, we manufactured S-I-S junctions from Nb/AlO$_x$/Nb
trilayer with a nominal critical current density of $j_c = 500\,$A/cm$^2$ (see
Ref.\:\cite{PTB-Nb} for details of the process). The upper Nb layer was coated
with a thin Al layer which was then oxidized forming the insulator for the
S-I-N sandwich. For the normal metal counter electrode we used a relatively
thick (100~nm) layer of AuPd alloy (normally used for thin-film on-chip
resistors). The specific tunnel resistance of the S-I-N structure was
$\rho\approx135\,\Omega\cdot\mu\textrm{m}^{2}$. The areas of S-I-S and S-I-N
junctions were 24$\,\mu$m$^2$ and 450$\,\mu$m$^2$, respectively. The
characteristics of the S-I-N sandwich are shown in upper inset of Fig.\,3. The
product $G_0R$ was found to be significantly smaller than expected which is
possibly due to imperfection in this rather transparent barrier. Nevertheless,
these values ($\approx 2.3$ at 4.2\,K increasing up to 6.2 at 1.4\,K and
yielding a fit value $\delta = 3.3$) still allowed rather nonlinear shunts to
be realized. The resulting non-hysteretic $I$-$V$ curves unambigously indicate
the overdamped regime of operation. The estimated values of $\beta_c$ are about
1. The curve measured at 4.2\,K is close to that given by the RSJ model, while
the shape of the curve at 1.4\,K is similar to that found in our model.

The $I$-$V$ curves in Fig.\,3 also show peculiarities due to the presence of an
additional S-I-S junction formed in the present technology directly beneath the
S-I-N sandwich and connected in series to it. Since its critical current was
sufficiently large, $I_{c}^{*} \approx 2.7\,\textrm{mA} \gg I_c$, it remained
in the superconducting state even at $I_c< I\leq I_{c}^{*}$. Its equivalent
circuit is displayed as parallel connection of Josephson inductance
$L^{*}=\Phi_0/2\pi I_c^{*}$ and capacitance $C^{*}$ (see lower inset in
Fig.\,3) yielding a plasma frequency $\nu_p=1/2\pi(L^{*}C^{*})^{1/2}\approx
70$\,GHz. Such a shunting path almost behaved as a sole S-I-N junction when the
Josephson oscillation frequency $2e\langle V\rangle/h$ was not close to
$\nu_p$. At $\langle V\rangle \approx h\nu_p/2e$ the impedance of the shunting
path resonantly increased and this enhanced (negative) contribution to the
average voltage due to self-detecting of Josephson oscillations. As a result,
the negative peaks with tips positioned at $\langle V\rangle \approx \pm
140\,\mu$V had emerged.

In conclusion, we have shown by modelling and preliminary experiment that
external shunting of S-I-S Josephson junctions by S-I-N junctions can ensure
both sufficiently high dynamic damping and low zero-voltage damping. For
operation at very low temperature ($\delta \gg 1$), the conditions
$\beta_c\lesssim 1$ and $\alpha \gtrsim 1$ lead to a limitation of the specific
resistance of S-I-N structure, viz. $\rho \lesssim \hbar/2c\Delta$, $c$ being
the specific capacitance of the barrier. For example, in the S-I-N structures
based on superconducting aluminum ($\Delta_{\textrm{Al}}\approx 200\,\mu$eV, $c
\approx 50\,\textrm{fF}/\mu\textrm{m}^2$) and operating at qubit (mK)
temperatures, $\rho$ should be $\lesssim 30\,\Omega\cdot\mu\textrm{m}^{2}$. The
manufacture of such sandwiches ensuring negligible zero-bias conductance
required for dramatic suppression of back action on the qubit is quite
feasible.

The authors would like to thank D.~Esteve, D.~Vion, M.\,G.~Castellano,
F.~Chiarello and K.~Arutyunov for stimulating discussions. This work was
partially supported by the EU through the SQUBIT-2 and RSFQubit projects.



\end{document}